# Photonic logic tensor computing beyond TOPS per core

Wenkai Zhang[1], Bo Wu[1], Wentao Gu[1], Hailong Zhou[1,*], Weida Hu[2,*], Ting He[3], Liao Chen[1], Wenchan Dong[1], Dongmei Huang[4], Yang Zhao[5], Wei Wang[5], Naidi Cui[5], Qiansheng Wang[6], Xi Xiao[6], Jianji Dong[1,*] and Xinliang Zhang[1,7]

[1]Wuhan National Laboratory for Optoelectronics, School of Optical and Electronic Information, Huazhong University of Science and Technology, Wuhan 430074, China.

[2]State Key Laboratory of Infrared Physics, Shanghai Institute of Technical Physics, Chinese Academy of Sciences, Shanghai, China.

[3]Hangzhou Institute for Advanced Study, University of Chinese Academy of Sciences, Hangzhou, China.

[4]Photonics Research Institute, Department of Electrical and Electronic Engineering, The Hong Kong Polytechnic University, Hong Kong SAR, China.

[5]United Microelectronics Center, Chongqing, 401332, China.

[6]National Information Optoelectronics Innovation Center, China Information and Communication Technologies Group Corporation, Wuhan 430074, China.

[7]Xidian University, 710071 Xi'an, China.

*Corresponding author: hailongzhou@hust.edu.cn; wdhu@mail.sitp.ac.cn; jjdong@hust.edu.cn;

**Abstract:** The soaring demand for computing resources has spurred great interest in photonic computing with higher speed and larger computing capacity. Photonic logic gates are of crucial importance due to the fundamental role of Boolean logic in modern digital computing systems. However, most photonic logic schemes struggle to exhibit the capability of massively parallel processing and flexible reconfiguration, owing to weak and fixed nonlinearity in optical elements. Here, we propose a photonic logic tensor computing architecture for the first time and fabricate the photonic universal logic tensor core (PULTC) with a parallel logic computing capacity beyond TOPS. Ten wavelength channels and four spatial channels are designed in PULTC, where the logic computing speed in each channel can reach 50 Gbit/s. After the nonlinear mapping of microring modulators, arbitrary logic operations can be achieved by configuring the Mach–Zehnder interferometer mesh. Our work offers an innovative route for photonic universal logic computing with high-parallel capability and propels the practical applications of photonic logic computing.

## Introduction

In this paper, we present a photonic logic tensor computing architecture that enables parallel processing of logic functions for the first time. By employing the EO nonlinearity of the microring modulators (MRMs), the electrical logic signals are nonlinearly mapped into a higher-dimensional vector space. Then, arbitrary logic operations can be directly achieved through linear transformation of the following Mach–Zehnder interferometer (MZI) mesh. The narrow bandwidth of MRMs and the broad bandwidth of the MZI mesh make it possible to multiplex massive wavelength channels, thereby performing logic operations in parallel. We fabricate the photonic universal logic tensor core (PULTC) with 10 independent wavelength channels and 4 independent spatial channels, where the EO bandwidth of MRMs is more than 50 GHz and the wavelength transmission bandwidth of the MZI mesh exceeds 80 nm. The total computing capacity exceeds TOPS per core. The PULTC fully leverages the advantages of optical parallelism and programming ability, paving the way for large-capacity universal photonic logic computing.

## Results

### Principle of PULTC

The novel photonic logic tensor computing architecture is shown in Fig. 1(a), which can support parallel logic computing with multiple input channels. Parallel electrical binary signals are first input into PULTC. Since two input binary signals will yield four possible combinations, all the output logic signals lie in a four-dimensional vector space. Therefore, it is not feasible to achieve arbitrary logic output through directly running a linear transformation of input Signal *A* and Signal *B*. Here, PULTC firstly maps the input electrical signals in a two-dimensional vector space to the output optical signals in a four-dimensional vector space by generating two linearly independent signals, i.e., continuous wave (*CW*) and Logic *AB* during the modulation process, where Logic *AB* is a nonlinear mapping process. Arbitrary logic output can be generated by programming the following optical linear network[1,2]. Considering each MRM operates at a certain resonance wavelength, multiple wavelength channels can be processed in parallel by using MRMs. The following optical linear network will execute the same logic functions for all input wavelength channels. We can obtain the corresponding logic results by demultiplexing the wavelength channels. Besides the wavelength dimension, the spatial dimension can also be utilized to simultaneously execute different logic operations for the same input signals. Parallel processing in both wavelength and spatial dimensions endows the PULTC with immense computing capacity.

The inner structure of PULTC is presented in Fig. 1(b), which is composed of a nonlinear mapping region and a linear transformation region. In the nonlinear mapping region, *CW* with multiple wavelengths is first divided equally into two channels. One channel remains *CW*, while the other is modulated by a series of MRMs. Here, we regard MRMs as optical switches. When the input electrical signal is Logic 0, the MRM is in the resonant state and the optical output is Logic 0. When the electrical signal is Logic 1, the MRM is in the non-resonant state and the optical output is Logic 1. The narrow bandwidth of the resonant peak allows each MRM to modulate one wavelength independently without mutual interference. Thus, the parallel input of electrical Signal *A* ($A_1 A_2 \ldots A_n$) can be loaded into the corresponding wavelength channels ($\lambda_1 \lambda_2 \ldots \lambda_n$). The *CW* and Signal *A* are then evenly split into two channels respectively. Among them, one channel of the *CW* and one channel of Signal *A* are directed into the cascaded dual-waveguide MRMs. Fig. 1(c) illustrates the MRM's modulation process of input electrical Signal *B*. With the electrical signal applied to the MRM, *CW* in the upper waveguide is modulated into Signal *B*, and Signal *A* in the lower waveguide undergoes the Logic AND operation with Signal *B*. Only when both Signal *A* and Signal *B* are Logic 1, the optical output of the lower waveguide is Logic 1. The other two channels retain the initial input signals, which output *CW* and Signal *A*, respectively. The nonlinear mapping process can be written as

$$E = \begin{bmatrix} A \\ B \end{bmatrix} \in S_E = \begin{bmatrix} 0 & 0 & 1 & 1 \\ 0 & 1 & 0 & 1 \end{bmatrix} \xrightarrow{nonlinear\ mapping} I = \begin{bmatrix} CW \\ B \\ AB \\ A \end{bmatrix} \in S_I = \begin{bmatrix} 1 & 1 & 1 & 1 \\ 0 & 1 & 0 & 1 \\ 0 & 0 & 0 & 1 \\ 0 & 0 & 1 & 1 \end{bmatrix}, \quad (1)$$

where $E$ and $I$ represent the input electrical signals and output optical signals of the nonlinear mapping region, $S_E$ and $S_I$ represent all the possible states of $E$ and $I$. The full rank of $S_I$ indicates that the input two-dimensional vector plane is mapped to a four-dimensional vector space, enabling the following part to achieve arbitrary logic operations through linear transformation. Note that this nonlinear mapping method is not limited to two input signals. It can also generate $2^N$ linearly independent vectors for *N*-input binary signals, which is discussed in **Supplementary 1**. Fig. 1(d) shows the linear transformation region composed of *m* columns of MZIs, where each column can perform 1×4 optical vector multiplication. We can configure the MZI mesh to perform different linear transformation by configuring the heaters on both outer and inner arms of MZIs. The linear transformation process is given by

$$O = \begin{bmatrix} T_1 \\ T_2 \\ \vdots \\ T_m \end{bmatrix} I = \begin{bmatrix} T_1 I \\ T_2 I \\ \vdots \\ T_m I \end{bmatrix}, \quad (2)$$

where $T_i$ represents the 1×4 linear transformation vector of *i*th column in the MZI mesh, $O$ represents the output optical logic matrix, and each row of $O$ represents the logic results ($T_i I$) at *i*th output port. Considering the optical signals $I$ can contain *n* wavelengths, the tensor transformation process can be expressed as

$$O = \begin{bmatrix} T_1 \\ T_2 \\ \vdots \\ T_m \end{bmatrix} [I_1, I_2, ..., I_n] = \begin{bmatrix} T_1 I_1 & T_1 I_2 & \cdots & T_1 I_n \\ T_2 I_1 & T_2 I_2 & \cdots & T_2 I_n \\ \vdots & \vdots & \ddots & \vdots \\ T_m I_1 & T_m I_2 & \cdots & T_m I_n \end{bmatrix} = \begin{bmatrix} O_{11} & O_{12} & \cdots & O_{1n} \\ O_{21} & O_{22} & \cdots & O_{2n} \\ \vdots & \vdots & \ddots & \vdots \\ O_{m1} & O_{m2} & \cdots & O_{mn} \end{bmatrix}, \quad (3)$$

where $I_j$ represents the output signal in the wavelength of $\lambda_j$ from the nonlinear mapping region and $O_{ij}$ represents the logic output from *i*th columns of the MZI mesh at wavelength of $\lambda_j$. By combining the spatial and wavelength dimensions, PULTC can execute *m*×*n* logic operations simultaneously, significantly enhancing the computing capacity by *m*×*n* times.

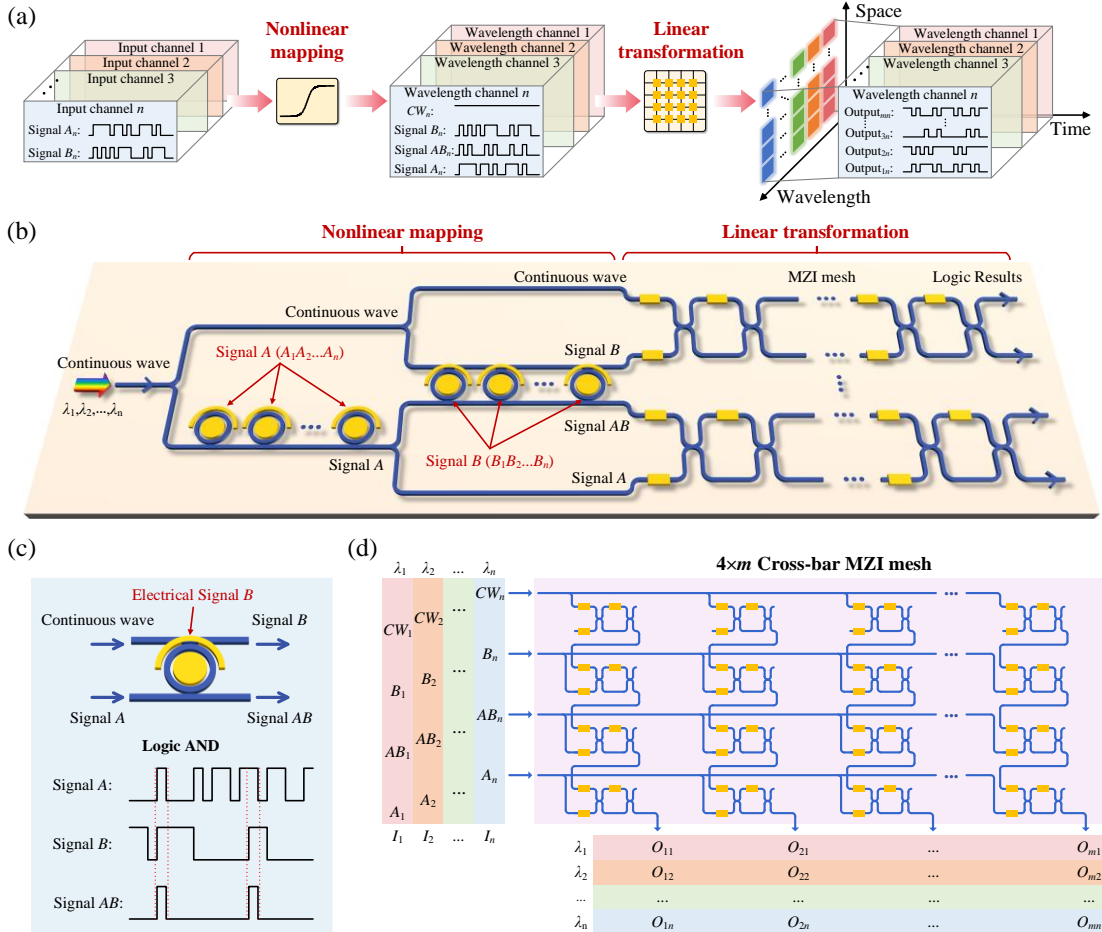

**Fig. 1 Operating principle of PULTC.** (a) Sketch map of photonic logic tensor computing. (b) Inner structure of PULTC, including

a nonlinear mapping region and a linear transformation region. (c) MRM's modulation process of input signal B. (d) 4×*m* linear transformation matrix realized by Cross-bar MZI mesh. *CW*, continuous wave.

## Demonstration of PULTC's functions

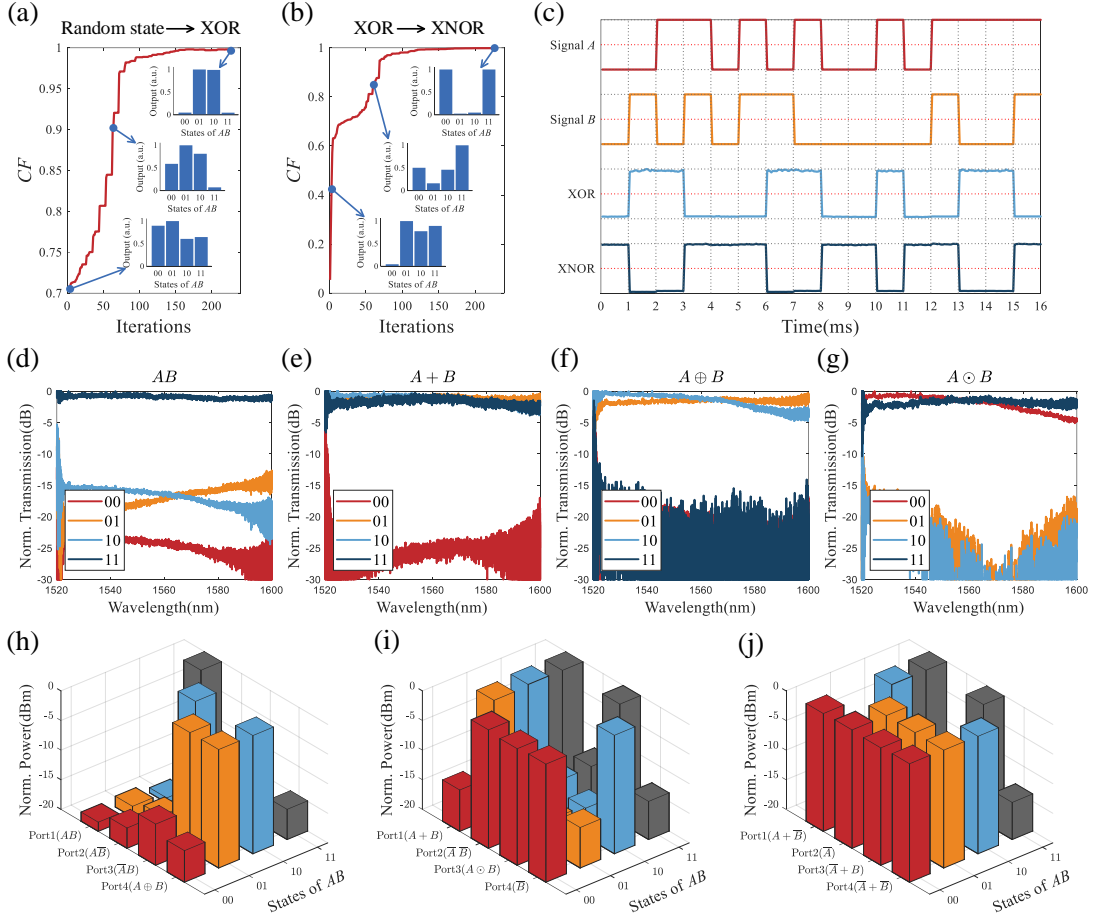

**Fig. 2 Basic performance measurement of PULTC.** Iteration processes of PULTC (a) from the random state to Logic XOR, and (b) from Logic XOR to Logic XNOR. (c) Waveforms of Logic XOR and Logic XNOR implemented by PULTC at the speed of 1 kbit/s. (d-g) Spectral responses of Logic AND, OR, XOR, and XNOR in the MZI mesh. (h-j) Parallel output of different logic results from four ports of PULTC.

We first select one output port in PULTC to verify the feasibility of logic functions. The MZI mesh is configured to execute the desired linear transformation according to the gradient descent algorithm[3]. The cost function *CF* is defined as

$$CF = \frac{\left|O^{tgt} \cdot O^{exp}\right|}{\left\|O^{tgt}\right\| \cdot \left\|O^{exp}\right\|} \quad (4)$$

where $O^{tgt}$ and $O^{exp}$ represent the vectors of the target logic and the experimental result, $|\ |$ is to get the absolute value, and $\|\ \|$ is 2-norm of the vector. All combinations of the two logic input signals (00, 01, 10, and 11) are explored in each iteration. The *CF* is the correlation between the target and experimental results, ranging from 0 to 1. Fig. 2(a) demonstrates the iteration process of PULTC from a random state to Logic XOR. As the *CF* gradually approaches 1, the output vector gets closer to target logic. The PULTC can easily switch to another logic function. By changing the target logic result $O^{tgt}$, the output logic will be updated from Logic XOR to Logic XNOR shown in Fig. 2(b). Once the voltage settings for each logic operation are obtained, the iteration process is no longer necessary. We can directly switch to the desired logic function by applying corresponding voltage settings to the MZI mesh. Logic output waveforms of XOR and XNOR in Fig. 2(c) reveal that the PULTC correctly performs logic functions. The extinction ratio (ER) of logic waveforms exceeds 13 dB, while the ER of MRM's resonant peak is less than 10 dB. This ER improvement indicates that the MZI mesh can enhance the signal quality to some extent during linear transformation of input signals. We then measure the transmission spectrum of the MZI mesh and calculate the spectral responses of four logic functions in Figs. 2(d-g) according to the theoretical input of the MZI mesh in PULTC. From 1520 nm to 1600 nm, all logic functions can reach the ER of over 10 dB. This bandwidth is sufficient to support the parallel operation of much larger than 10 wavelengths. Figs. 2(h-j) depict the parallel output of different logic results from four spatial ports of PULTC under three different configurations. The ER of all logic functions can exceed 13 dB, illustrating the excellent performance of PULTC's parallel processing in the spatial dimension.

Fig. 3(a) depicts the high-speed test platform for PULTC, where electrical Signal *A* and Signal *B* are applied to MRMs' GSG pads by microwave probes. To facilitate the control of PULTC, we conduct optoelectronic packaging on the chip. The fiber arrays couple lightwave into and out from the chip's optical gratings. Through wire bonding, the voltages of heaters in MRMs and MZIs are applied by a multi-channel power source. We utilize the thermo-electric cooler to sustain the environmental temperature of PUTLC. The measured EO bandwidth of the MRM in Fig. 3(b) exceeds 50 GHz, demonstrating its satisfactory modulation performance. By configuring the MZI mesh, we realize 14 different logic functions at the bit rate of 25 Gbit/s, which cover all possible cases of two binary signals except for the outputs of all zeros and all ones, as shown in Figs. 3(c). The 14 different logic results are consistent with the theoretical calculations, indicating the universality and reconfigurability of PULTC.

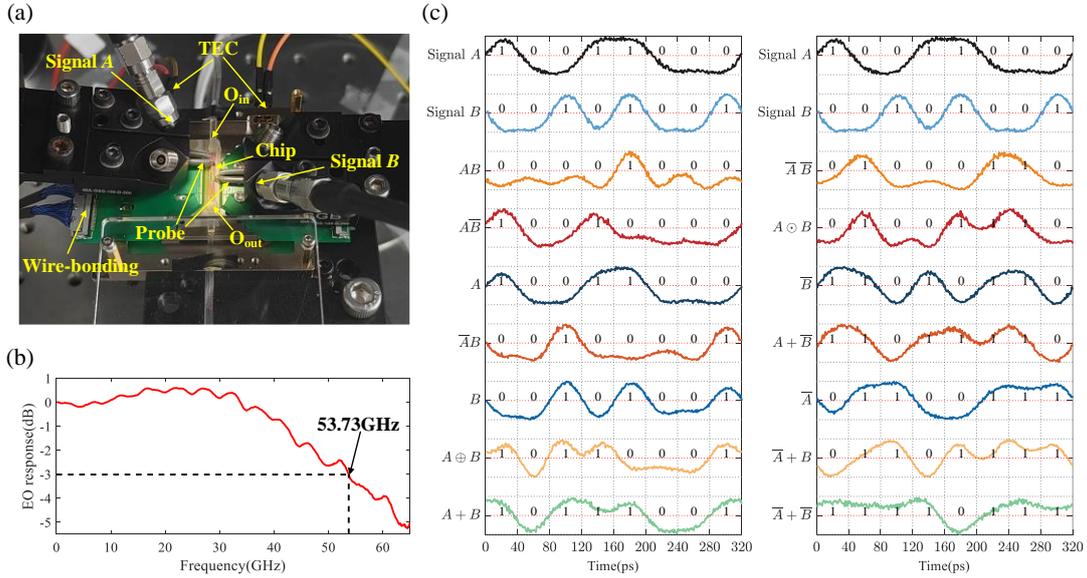

**Fig. 3 High-speed performance characterization of PULTC.** (a) High-speed test platform for PULTC. (b) EO response of the MRM. (c) Waveforms of input logic signals and 14 different logic operations at 25Gbit/s. TEC, thermo-electric cooler;

## Conclusion

In summary, we devise a photonic logic tensor computing architecture to realize parallel logic computing for the first time. The PULTC is fabricated that supports 10 wavelength channels and 4 spatial channels. The logic computing speed in one single channel can reach 50 Gbit/s, enabling the total parallel computing capacity of PULTC beyond TOPS per core. After optimization, the computing capacity can reach 40 TOPS. The proposed PULTC opens a new avenue for large-capacity, high-speed, high-efficiency universal photonic logic computing.